# Tıbbi Dokümanların Ayrıştırılmasında Kullanılan Sınıflandırma Algoritmalarının Karşılaştırılması

# Comparison of Classification Algorithms Used Medical Documents Categorization

Durmuş Özkan ŞAHİN[1], Erdal KILIÇ[1]
[1]Bilgisayar Mühendisliği Bölümü, Ondokuz Mayıs Üniversitesi, Samsun, Türkiye
{durmus.sahin, erdal.kilic}@bil.omu.edu.tr

*Özetçe*— Metin tabanlı belgelerin hacmi günden güne artmaktadır. Tıbbi dokümanlarda bu artan metin belgelerinin içinde yer almaktadır. Bu çalışmada metin sınıflandırma için kullanılan teknikler tıbbi dokümanlar üzerinde uygulanmış ve sınıflandırma başarısı değerlendirilmiştir. Kullanılan veri setleri çok sınıflı ve çok etiketlidir. Öznitelik seçme yöntemi için Chi Square (CHI) tekniği ayrıca sınıflandırma için SMO, NB, C4.5, RF ve KNN algoritmaları kullanılmıştır. Çalışmanın amacı, çok sınıflı ve çok etiketli tıbbi dokümanlardan oluşmuş veri seti üzerinde çeşitli sınıflandırıcıların başarıları değerlendirilmiştir. İlk 400 öznitelikte KNN en başarılı sınıflandırıcı olurken, öznitelik sayısı 400 ve sonrasında SMO en başarılı sınıflandırıcı olmuştur.

*Anahtar Kelimeler — metin sınıflandırma; metin madenciliği; tıbbi dokümanlar.*

*Abstract*— Volume of text based documents have been increasing day by day. Medical documents are located within this growing text documents. In this study, the techniques used for text classification applied on medical documents and evaluated classification performance. Used data sets are multi class and multi labelled. Chi Square (CHI) technique was used for feature selection also SMO, NB, C4.5, RF and KNN algorithms was used for classification. The aim of this study, success of various classifiers is evaluated on multi class and multi label data sets consisting of medical documents. The first 400 features, while the most successful in the KNN classifier, feature number 400 and after the SMO has become the most successful classifier.

## I. GİRİŞ

İnternetin yaygınlaşması ile birlikte bilgisayarların, cep telefonlarının ve tabletlerin kullanımı her geçen gün artmaktadır. Bu artış ile sosyal ağların, elektronik kütüphane, elektronik kitap ve elektronik posta gibi metin biçiminde üretilen ve depolanan verilerin sayısında kayda değer bir artış görülmektedir. Bu durumun sonucunda metin içeren verilerin sayısı bir hayli artmış ve artmaya devam edecektir. Bu nedenle bu verileri bilgisayarlar vasıtasıyla otomatik olarak işlemek, bu verilerden anlamlı bilgiler elde edip kullanmak insanların işini kolaylaştıracaktır.

Tıbbi dokümanların sınıflandırılması, analiz edilmesi de bu problemler arasında yer almaktadır. Tıbbi dokümanlar üzerinde birçok çalışma yapılmıştır. Spat ve arkadaşları Almanca tıbbi dokümanların sınıflandırılması üzerine bir çalışma yaparak sınıflandırma algoritmalarının performansını karşılaştırmışlardır [1]. Ancak bu çalışmanın eksik yanı hangi öznitelik seçme yönteminin kullanıldığının belli olmamasıdır. Oğuz yaptığı yüksek lisans tez çalışmasında kulak-burun-boğaz polikliniğinde tedavi görmüş hastalara ait bilgi formlarının yapısal hale getirilmesini sağlamıştır [2]. Bu dokümanlar üzerinde birliktelik analizi kurallarını uygulayarak hekimlerin hasta ile ilgili ihtiyaç duydukları bilgilere erişiminin kolaylaştırılması hedeflenmiştir. MeSH (Medical Subject Heading) tıp, hemşirelik, diş hekimliği ve veteriner sağlığı gibi konular ile ilişkilendirilmiş kontrollü bir anahtar kelime veri tabanıdır [3]. MeSH veri tabanında yer alan çeşitli hastalıklara ait ilgili kelimeler, doküman içerisinde geçip geçmemesine göre sınıflandırılmaktadır. Elberrichi ve arkadaşları hem MeSH veri tabanından hem de dilin anlamsal özelliklerinden yararlanarak dokümanları sınıflandırmıştır [4]. Kelimeler arasında ki anlamsal özellikler ise hyperonyms tekniği ile saptanmıştır [4]. Lopez ve arkadaşları ise hem istatistiksel hem de anlamsal özelliklere bakarak öznitelik seçimi yapmışlar, Naïve Bayes (NB) ve Rocchio sınıflandırma algoritmaları ile dokümanları sınıflandırmışlardır [5]. Kelimelerin anlamsal ilişkileri UMLS (Unified Medical Language System) sisteminden elde edilerek belirlenmektedir. Bu sistemde tıbbi bazı kelimelerin eş anlamları yer almaktadır [5]. Kocatekin ve arkadaşı metin madenciliği tekniklerini kullanarak Türkçe radyoloji raporlarının sınıflandırılmasını yapmışlardır. Yapılan çalışmada öznitelik seçimi elle yapılarak sınıflandırma gerçekleştirilmiştir [6]. Parlak ve arkadaşı MEDLINE veri tabanında yer alan en çok dokümana sahip 10 hastalık sınıflı tıbbi makaleleri alarak çok sınıflı tek etiketli bir veri seti oluşturmuşlardır [7]. Bu veri seti üzerinde bir ön işlem adımı olan kök bulmanın sınıflandırma algoritmaları üzerinde performansları karşılaştırılmıştır. Çalışma da metin sınıflandırma da sıklıkla kullanılan öznitelik seçme yöntemleri kullanılmamış fakat boyut indirgeme amacıyla bir kelime 5 den az dokümanda geçiyorsa o kelime elenerek öznitelik olarak kabul edilmemiştir [7]. Haltaş ve arkadaşı ise MEDLINE veri tabanında yer alan tıbbi dokümanların kanser türlerine göre sınıflandırılmasını yapmıştır [8]. Hem CHI hem de Information Gain (IG) öznitelik seçme yöntemleri kullanılarak sınıflandırma algoritmalarının performansları karşılaştırılmıştır [8].

Bu çalışmada ise MEDLINE veri tabanının alt kümesi olan OHSUME veri seti kullanılmıştır. Bu veri setinin özelliği ise çok sınıflı ve çok etiketli olmasıdır. Bu veri seti üzerinde CHI ile öznitelik seçimi yapılarak çeşitli sınıflandırma algoritmalarının performansları karşılaştırılmıştır. Çalışmanın asıl amacı çok sınıflı ve çok etiketli tıbbi dokümanlardan oluşmuş veri seti üzerinde çeşitli sınıflandırıcıların sınıflandırma başarıları değerlendirilecektir. Çalışmanın diğer çalışmalardan farkı kullanılan veri setinin çok sınıflı ve çok etiketli olmasının yanında farklı çalışma sistemlerine sahip çok sayıda sınıflandırıcının karşılaştırılmasıdır. Ayrıca hesaplama maliyeti çok fazla olmayan ve metin sınıflandırmada sıkça kullanılan CHI metriği ile kategorilere ait ayırt edici terimlerin getirilmesi sağlanmıştır.

## II. MATERYAL ve YÖNTEM

Metinler HTML, PDF, DOC gibi bilgisayarların okuyabileceği dosyalarda tutulmasına rağmen sınıflandırma algoritmalarının anlayabileceği yapılarda değillerdir. Bu nedenle bir takım işlemlerden geçirilerek yapısal bir hale getirilmelidir. Metinleri yapısal hale getirip sınıflandırmak için ön işlem, terim ağırlıklandırma, öznitelik seçimi ve sınıflandırma işlemleri uygulanır. Bu bölümde bu adımlar detaylı olarak anlatılacaktır.

### A. Ön İşlem

Dokümanlar içerisinde yer alan kelimeler her zaman aynı formda değildir. Bu kelimeler aynı öneme sahip olduğu için aynı biçime dönüştürülmelidir. Aynı biçime dönüştürülmezse kelimelerin geçiş sayıları farklı olacağından yanıltıcı sonuçlar alınmasına neden olabilir. Bu çalışmada bütün kelimelerin karakterleri büyük harften küçük harfe dönüştürülmüş, noktalama işaretleri ve rakamlar silinmiş, bütün kelimeleri elde etmek için boşluğa göre parçalama yapılmış (tokenization), her kelimenin kökü bulunmuş ve dilde sıkça geçen durak kelimeler silinmiştir. Kelimelerin kökünü bulmak için İngilizce kök bulma algoritması Porter Stemmer kullanılmıştır [9].

### B. Terim Ağırlıklandırma

Metin madenciliğinde terim ağırlıklandırma yapılmasının en önemli sebebi sınıflandırma algoritmalarının ayırt edici gücünü daha belirgin şekilde ortaya çıkartmaktır. Örneğin herhangi bir terim bazı dokümanlarda çok az, bazı dokümanlarda çok fazla görünüyorsa bu farkın terim ağırlıklandırma ile ortaya çıkartılması gerekir. Bu çalışmada metin sınıflandırmada çok sık kullanılan TF.IDF (Term Frequency – Inverse Document Frequency) terim ağırlıklandırma yöntemi kullanılmıştır. TF ilgili terimin dokümanda geçiş adedi iken, IDF ise

$$IDF = \log(\frac{N}{a+c}) \quad (1)$$

Denklem (1)' de verilmiştir. Burada N, toplam doküman sayısı, a ilgili terimin pozitif kategorilerde geçtiği dokümanların sayısı, c ise ilgili terimin negatif kategorilerde geçtiği dokümanların sayısıdır. Terimin ağırlığı, TF ile IDF değerinin çarpılmasıyla hesaplanır.

### C. Öznitelik Seçimi

Sınıflandırma algoritmaları yapıları gereği çalışması uzun zaman almaktadır. Aynı zamanda bu algoritmaların çalışması verinin boyutu ile doğrudan ilişkilidir. Geleneksel metin sınıflandırma yaklaşımlarında kelime çantası (bag of words) tercih edilmektedir. Burada eğitim dokümanları içerisinde yer alan bütün kelimeler kullanılarak doküman vektörleri oluşturulur. Bu nedenle dokümanlar on binlerce terimlerle temsil edilirler. Bu terimlerin büyük bir çoğunluğu sınıflandırma başarısını olumlu yönde etkilemediği için önemli bilgiler içermediği düşünülür. Aynı zamanda vektör boyutu çok büyük olacağı için de bellek ve hesaplama problemi ortaya çıkacaktır.

Öznitelik seçimi, kelime çantasında yer alan tüm terimleri kullanmak yerine, tüm kelimeleri en iyi temsil eden terimlerin seçilerek, daha küçük boyutlu vektör elde edilmesi işlemidir. Yani öznitelik ile oluşturulmuş vektör, kelime çantasının en iyi alt kümesidir. Öznitelik seçimi sayesinde,

- Vektör boyutu azaltılarak bellek israfı önlenebilir.
- Metin sınıflandırma işleminin çalışma zamanı azaltılabilir.
- Önemli bilgiler içermediği düşünülen veya gürültü olarak kabul edilen veriler ortadan kaldırılabilir.

Öznitelik seçme yöntemleri genellikle 3 ana gruba ayrılır [10, 11]. Bunlar filtrelemeli (filter), sarmalamalı (wrapper), gömülü (embedded) metotlardır. Bu yöntemlerin yanında karma yöntemlerde mevcuttur [10]. Bu çalışmada filtrelemeli yöntemlerden olan CHI ile öznitelik seçimi yapılmıştır. CHI için;

$$CHI(t_j, c_i) = N \frac{(ad-bc)^2}{(a+c)(b+d)(a+b)(c+d)} \quad (2)$$

Denklem (2) verilmiştir. Burada $c_i$ sınıfında yer alan $t_j$ teriminin CHI skoru hesaplanmaktadır. Denklem (2)' de a $t_j$ teriminin pozitif kategorilerde geçtiği dokümanların sayısı, b $t_j$ teriminin pozitif kategorilerde geçmediği dokümanların sayısı, c $t_j$ terimin negatif kategorilerde geçtiği dokümanların sayısı, d $t_j$ terimin negatif kategorilerde geçmediği dokümanların sayısı, N ise toplam doküman sayısıdır. Bütün terimlerin CHI skorları hesaplanır. Bu skorlar büyükten küçüğe doğru sıralanarak kategori ile en ilişkili kelimeler çıkartılmış olur. CHI skoru 0 ise kategori ile terim arasında hiçbir ilişki yoktur.

*D. Sınıflandırma*

Sınıflandırma bir denetimli öğrenme problemidir. Etiketleri belli olan metinler önce eğitilir, eğitim aşamasından sonra yeni metinlerin sınıflandırıcının modeline göre hangi kategoriye ait olduğu belirlenir. Metin sınıflandırma çalışmalarında genellikle ikili sınıflandırma tercih edilir. Sınıflandırılacak olan kategori pozitif olarak etiketlendirilirken, diğer kategorilerin hepsi negatif olarak etiketlenmektedir. Bu çalışmada ikili sınıflandırma tekniği uygulanmıştır. İkili sınıflandırma kullanılmasının amacı kullanılan veri setinin çok sınıflı ve çok etiketli olmasından dolayıdır. Aksi durumda bazı metinler birden fazla kategoride yer almasına rağmen yalnızca bir kategori için sınıflandırılıp yanlış sonuç alınacaktır.

Çalışmanın sınıflandırma aşamasında WEKA aracı kullanılmıştır [12]. WEKA aracı içerisinde sınıflandırma algoritmalarından kümeleme algoritmalarına kadar birçok veri madenciliği yöntemi yer almaktadır. Geliştirilen uygulamada metin sınıflandırma alanında oldukça popüler olan doğrusal ayrılabilme ilkesine dayanan Support Vector Machine (SVM), olasılık tabanlı NB, karar ağaçlarından olan C4.5 ve Random Forest (RF), örnek tabanlı bir sınıflandırma algoritması olan K-Nearest Neighbors (KNN) kullanılmıştır. WEKA aracı içerisinde SVM doğrudan olmazken, SVM kullanan Sequential Minimal Optimization (SMO) algoritması ile sonuç alınmıştır.

*E. Kullanılan Veri Seti*

Bu çalışmada MEDLINE veri tabanının alt kümesi olan OHSUME veri seti alınarak tıbbi dokümanlar sınıflandırılmaya çalışılmıştır. Veri setinde yer alan çeşitli hastalıklara ait dokümanların dağılımı Tablo 1'de verilmiştir.

| Sınıf | Eğitim | Test |
|---|---|---|
| Bacterial infections and mycoses | 423 | 506 |
| Virus diseases | 158 | 233 |
| Parasitic diseases | 65 | 70 |
| Neoplasms | 1163 | 1467 |
| Musculoskeletal diseases | 283 | 429 |
| Digestive system diseases | 588 | 632 |
| Stomatognathic diseases | 100 | 146 |
| Respiratory tract diseases | 473 | 600 |
| Otorhinolaryngologic diseases | 125 | 129 |
| Nervous system diseases | 621 | 941 |

**Tablo 1.** Kullanılan veri seti ve hastalıklara göre dağılımı

*F. Performans Ölçütü*

Performans ölçeği ile belgenin ilgili sınıfa ait olma doğruluğu ölçülür. Veri setinde gerçekte pozitif etiketli olan bir örnek, sınıflandırma sonucunda pozitif olarak sınıflandırılıyorsa True Positive (TP) olarak adlandırılır. Gerçekte negatif etiketli olan bir örnek, sınıflandırma sonucunda negatif olarak sınıflandırılıyorsa True Negative (TN) olarak adlandırılır. Gerçekte negatif olan bir örnek pozitif olarak sınıflandırılıyorsa False Positive (FP) ve gerçekte pozitif olan bir örnek negatif olarak sınıflandırılıyorsa False Negative (FN) olarak isimlendirilir. Tablo 2'de hata matrisi (confusion matrix) şeklinde bu durumlar gösterilmektedir.

| Gerçek / Tahmin | C1 | C2 |
|---|---|---|
| C1 | TP | FP |
| C2 | FN | TN |

**Tablo 2.** Hata Matrisi (Confusion Matrix)

Metin sınıflandırmada en çok kullanılan performans ölçeği F-ölçütüdür. F-ölçeği duyarlılık (precision) ve anma (recall) değerlerinin harmonik ortalamasıdır. Denklem (3)'de duyarlılık ($\pi$) değeri, Denklem (4)'de anma ($\rho$) değeri verilmektedir.

$$\pi = \frac{TP}{TP+FP} \qquad (3)$$

$$\rho = \frac{TP}{TP+FN} \qquad (4)$$

F-ölçütü ise Denklem (5) ile hesaplanmaktadır.

$$F = \frac{2\pi\rho}{\pi+\rho} \qquad (5)$$

Denklem (5) $F_1$ veya $F_{micro}$ olarak da bilinmektedir. Bu çalışmada sınıflandırma başarımını ölçmek için $F_{micro}$ kullanılmıştır.

## III. SONUÇLAR

Farklı çalışma modeline sahip 5 farklı sınıflandırıcıların, tıbbi dokümanlar üzerinde başarım sonuçları Şekil 1'de verilmiştir.

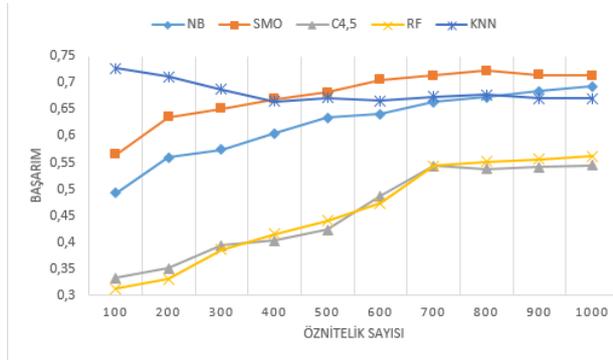

**Şekil 1.** Sınıflandırma algoritmalarının başarımı

Sınıflandırma algoritmalarının performansları arasında çok belirgin fark olmakla birlikte en başarılı sınıflandırma sonucu 100 öznitelikte ( $F_{micro} = 0,7275$ ) KNN algoritması ile elde edilmiştir. KNN algoritması için k değeri 1 alınarak sınıflandırma yapılmıştır. KNN algoritmasının olumsuz yanı ise öznitelik sayısı (veri boyutu) arttıkça sınıflandırma başarımı genelde azalmaktadır. Diğer sınıflandırma algoritmalarında ise öznitelik sayısı arttıkça sınıflandırıcıların başarımında bir artış meydana gelmektedir. SMO algoritması ile 800 öznitelik kullanılarak yapılan sınıflandırma da ( $F_{micro} = 0,7228$ ) ile ikinci en yüksek sınıflandırma başarısı elde edilmiştir. İlk 400 özniteliğe kadar KNN algoritması en iyi sonucu verirken, 400 öznitelik ve sonrasında ise SMO algoritması en iyi sonucu vermektedir. NB, C4.5 ve RF algoritmaları ise diğer sınıflandırıcıların gerisinde kalmaktadır. Karar ağacı tabanlı çalışan C4.5 ve RF az öznitelikte en kötü sonuçları vermesine rağmen öznitelik sayısı arttıkça performanslarında iyi bir artış görülmüştür. Fakat NB algoritması kadar başarılı olmadığı gözlenmektedir.

## IV. GELECEK ÇALIŞMALAR

Yapılan bu çalışmada çeşitli hastalıklara ait, çok sınıflı ve çok etiketli İngilizce tıbbi dokümanlar farklı çalışma modellerine sahip sınıflandırıcılar tarafından sınıflandırılarak sonuçlar değerlendirilmiştir. Ayrıca öznitelik seçiminde sıklıkla kullanılan CHI testi ile tıbbi dokümanlar üzerinde öznitelik seçimi yapılmıştır. Çeşitli hastalıklara ait Türkçe tıbbi dokümanların elde edilmesi halinde dokümanların otomatik bir şekilde sınıflandırılması mümkün olacaktır. Ayrıca bu çalışmada tıbbi olmayan dokümanların sınıflandırılmasında kullanılan durak kelimeler listesi kullanılmıştır. Bu listeye tıbbi dokümanlar üzerinde ayırt edici özelliği olmayan "disease" gibi kelimeler eklenerek daha iyi bir kelime seçimi yapılabilir.